\documentclass[letter]{aa}  
\usepackage{graphicx}
\usepackage{amsmath}
\usepackage{amsfonts}
\usepackage{amssymb}
\usepackage{gensymb}
\usepackage{textcomp}
\usepackage[varg]{txfonts}
\usepackage{natbib}
\bibpunct{(}{)}{;}{a}{}{,}
\begin{document}  


\title{Nature of the energy source powering solar coronal loops driven by nanoflares}

   \author{Authors}
   \author{L.~P.~Chitta\inst{1}, H.~Peter\inst{1}, \and S.~K.~Solanki\inst{1,}\inst{2}}

   \institute{Max Planck Institute for Solar System Research, 37077, G\"{o}ttingen, Germany\\
              \email{chitta@mps.mpg.de} 
              \and 
              School of Space Research, Kyung Hee University, Yongin, Gyeonggi, 446-701, Republic of Korea}

   \date{Received 10 May 2018; accepted 28 June 2018}

\abstract
{Magnetic energy is required to heat the
corona, the outer atmosphere of the Sun, to millions of degrees.}
%
{We study the nature of the magnetic energy source that is probably responsible for the brightening of coronal loops driven by nanoflares in the cores of solar active regions.}
%
{We consider observations of two active regions (ARs), 11890 and 12234, in which nanoflares have been detected. To this end, we use ultraviolet (UV) and extreme ultraviolet (EUV) images from the Atmospheric Imaging Assembly (AIA) onboard the Solar Dynamics Observatory (SDO) for coronal loop diagnostics. These images are combined with the co-temporal line-of-sight magnetic field maps from the Helioseismic and Magnetic Imager (HMI) onboard SDO to investigate the connection between coronal loops and their magnetic roots in the photosphere.}
%
{The core of these ARs exhibit loop brightening in multiple EUV channels of AIA, particularly in its 9.4\,nm\ filter. The HMI magnetic field maps reveal the presence of a complex mixed polarity magnetic field distribution at the base of these loops. We detect the cancellation of photospheric magnetic flux at these locations at a rate of about $10^{15}$\,Mx\,s$^{-1}$. The associated compact coronal brightenings directly above the cancelling magnetic features are indicative of plasma heating due to chromospheric magnetic reconnection.}
%
{We suggest that the complex magnetic topology and the evolution of magnetic field, such as flux cancellation in the photosphere and the resulting chromospheric reconnection, can play an important role in energizing active region coronal loops driven by nanoflares. Our estimate of magnetic energy release during flux cancellation in the quiet Sun suggests that chromospheric reconnection can also power the quiet corona.}


   \titlerunning{Energy source powering solar coronal loops driven by nanoflares}
   \authorrunning{L. P. Chitta et al.}
   \maketitle


\section{Introduction}\label{sec:intro}  

The extreme ultraviolet (EUV) and the thermal X-ray emission from the corona of the Sun is radiated by plasma at temperatures in excess of 1\,MK, which is much hotter than the Sun's surface temperature of about 6,000\,K. Spatially resolved solar observations reveal that, at EUV and X-ray wavelengths, the corona is particularly bright in active regions (ARs) hosting strong magnetic field concentrations including sunspots. The cores of these ARs have an emission component from hot plasma at temperatures of about 5\,MK \citep{2012ApJ...750L..10T} confined by the magnetic field structured in the form of compact loops. The origin of such high-temperature plasma and, more generally, the nature of the energy source responsible for coronal heating remain poorly understood. Any heating process has to explain not only the high temperature, but also the observed spatial structuring and temporal intermittency in the corona \citep[e.g.][]{1981SoPh...69...99W,1992PASJ...44L.147S,1997ApJ...482..519F,2014ApJ...783...12U}.

During the emergence of an AR, when new sunspots form, the energy released from the rapid reconfiguration of surface magnetic fields, including emergence and cancellation of magnetic flux, can heat the plasma to several million Kelvin \citep[e.g.][]{2011ApJ...726...12E}. However, the active corona persists for several days or even weeks, well after the emergence of large-scale magnetic field has ceased \citep[e.g.][]{2014ApJ...783...12U}. Therefore, current scenarios of coronal heating rely on either the dissipation of magnetohydrodynamic (MHD) waves \citep[e.g.][]{1981SoPh...70...25H,2015ApJ...807..146A,2017A&A...604A.130K}, or Ohmic dissipation of current sheets in nanoflares, induced by slowly moving the footpoints of coronal loops and braiding the magnetic field \citep[e.g.][]{1988ApJ...330..474P,2002ApJ...576..533P,2006SoPh..234...41K}. MHD waves are observed to be ubiquitous in the solar atmosphere \citep{2007Sci...317.1192T} and they can heat the quiet corona, i.e. the corona outside ARs \citep{2011Natur.475..477M}, but in general the observed wave power is two-orders-of-magnitude too weak to power the AR corona \citep{1977ARA&A..15..363W}. Although coronal loops may develop ubiquitous magnetic braids due to continual motion of their footpoints at the surface, direct observations of magnetic braiding, which would indicate the release of energy through unwinding of those braids, are sparse \citep{2013Natur.493..501C}. Moreover, questions were raised on the kind of footpoint motions that would generate such highly braided structures \citep{2014ApJ...787...87V}. Furthermore, \citet{2014ApJ...795L..24T} observed that the coronal brightenings at these braids were not triggered internally, but were initiated externally by photospheric flux cancellation along a polarity inversion line over which the braided structure had formed.

Recent studies that connect the evolution of coronal brightenings to the evolution of photospheric magnetic field in ARs find that some of the bright coronal loops are rooted in regions of mixed magnetic polarities \citep[i.e. regions where a dominant magnetic polarity and minor opposite magnetic polarity patches are present;][]{2017ApJS..229....4C,2017ApJ...843L..20T,2018ApJ...853L..26H}. Interactions of such mixed-polarity magnetic fields leading to the cancellation of surface magnetic flux followed by disturbances in the solar atmosphere that are widely associated with magnetic reconnection (e.g. plasma jets and compact brightenings) are also observed at the footpoints of coronal loops \citep{2017ApJS..229....4C,2017A&A...605A..49C,2018ApJ...853L..26H}. 

Based on the above observations, we find it necessary to investigate the nature of the magnetic energy source that is likely responsible for powering the coronal loops, in particular loops in the cores of ARs hosting hot plasma at several million Kelvin. In the present work, we consider brightenings in such loops that are associated with nanoflares. We find that these loops are apparently rooted in photospheric regions with a complex magnetic landscape containing mixed magnetic polarities. We notice that intermittent brightenings in the solar atmosphere follow surface magnetic flux cancellation at these locations. Our findings hint at the possibility of reconnection driven by the cancellation of surface magnetic flux as the energy source, at least in some of the coronal loop brightenings.

\section{Observations}\label{sec:obs}   

To investigate the origin of coronal loop brightenings in AR cores, we consider observations of AR 11890 and AR 12234 obtained with the Solar Dynamics Observatory \citep[SDO;][]{2012SoPh..275....3P}.
We use time series of images recorded in the ultraviolet (UV) at a cadence of 24\,s\ and the EUV at a \ cadence of 12\,s by the Atmospheric Imaging Assembly \citep[AIA;][]{2012SoPh..275...17L} onboard SDO to detect brightenings through the solar atmosphere. To trace the magnetic roots of these coronal loops at the solar surface, we analyse co-temporal, photospheric line-of-sight magnetic field maps of these ARs, recorded every 45\,s\ by the Helioseismic and Magnetic Imager \citep[HMI;][]{2012SoPh..275..207S} onboard SDO.

As described at the end of Sect.\,\ref{sec:intro}, these ARs are chosen simply because the loop brightenings in their cores are attributed to nanoflares. For instance, \citet{2014Sci...346B.315T} presented imaging and spectroscopic observations of AR 11890, where they detect footpoint brightenings at the base of a coronal loop that is visible in the AIA 9.4\,nm\ filter. These footpoint brightenings are associated with modest upflows with velocities of 15\,km\,s$^{-1}$ in the transition-region Si\,{\sc iv}\ observations obtained from the Interface Region Imaging Spectrograph. Based on numerical modelling of impulsively heated loops, the authors argued that the footpoint brightening and the upflows in the transition region are due to the interaction of the lower atmosphere with non-thermal electrons accelerated in coronal nanoflares. Similarly, \citet{2017NatAs...1..771I} presented observations from the Focusing Optics X-ray Solar Imager, with a detection of faint emission in hard X-rays (above 3\,keV) in AR 12234 that they ascribed to plasma heated above 10\,MK in nanoflares.

\section{Brightening in AR 11890}\label{sec:ar1}    


\begin{figure*}
\begin{center}
\includegraphics[width=\textwidth]{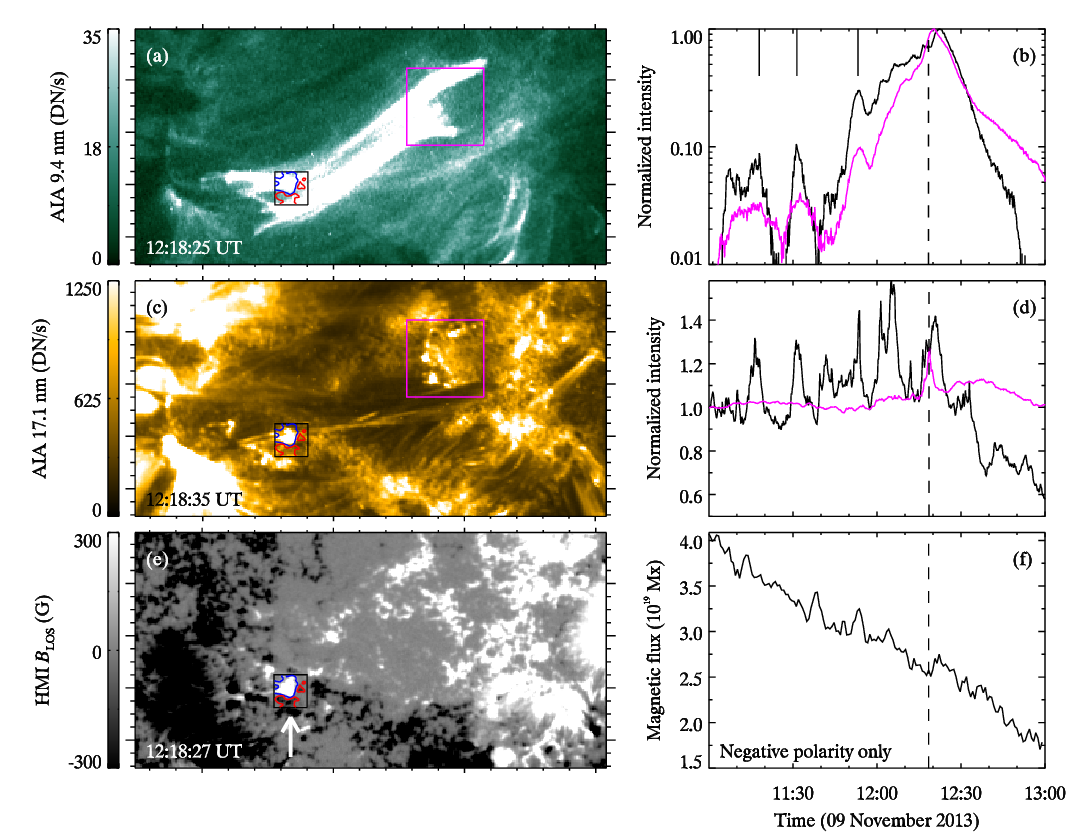}
\caption{Hot core loop in AR 11890 and its association with the cancellation of surface magnetic flux. (a) AIA 9.4 nm filter image showing a fully formed coronal loop seen from almost directly above. The black (east) and magenta (west) boxes mark the regions near the footpoints. The field of view is $180\arcsec\times90\arcsec$. (b) AIA 9.4 nm light curves from the footpoint regions. The short vertical bars point to the near-simultaneous peaks in the two light curves. (c) and (d) as in (a) and (b) but with the AIA 17.1 nm filter. (e) HMI line-of-sight magnetic field map, in which darker shading represents negative polarity and lighter shading positive polarity field. The black box encloses magnetic flux near the eastern footpoint of the loop; it lies at the same position as the black boxes in panels (a) and (c). The blue (positive polarity) and red (negative polarity) contours within this box cover regions with a magnetic flux density of $\pm75$\,G\ or more (see also panels a and c). (f) Magnetic flux of the negative polarity near the eastern footpoint as a function of time (see arrow). The vertical dashed line in the right panels marks the time-stamp of the corresponding spatial maps displayed on the left. See online movie to follow the evolution of this loop brightening. North is up. See Sect.\,\ref{sec:ar1} for details.
\label{fig:f1}}
\end{center}
\end{figure*}

Here we describe the case of a coronal loop brightening in AR 11890 observed on 9 November 2013 (Fig.\,\ref{fig:f1}); \citet{2014Sci...346B.315T} presented observations covering western footpoints of this loop (including the region roughly covered by the magenta box in Fig.\,\ref{fig:f1}) during the same period. This loop brightening is clearly distinguishable in the 9.4\,nm\ filter of AIA for about 40\,minutes. This filter can detect plasma at temperatures of about 7\,MK, but is also sensitive to cooler plasma at temperatures of about 1\,MK. During the phase when the main section of the loop is bright in the 9.4\,nm\ filter (Fig.\,\ref{fig:f1}(a)), it was invisible (or indistinguishable) to the other filters of AIA at 17.1\,nm, 19.3\,nm, or 21.1\,nm, that are more sensitive to plasma at $\sim$1\,MK\ \citep{2012SoPh..275...41B}, which suggests that most of the brightening  in the 9.4 nm image is due to hot plasma. This is demonstrated using the 17.1\,nm\ filter image displayed in Fig.\,\ref{fig:f1}(c). 

The loop brightening presented some interesting characteristics. Once the hot core loop is fully developed (Fig.\,\ref{fig:f1}(a)), it connects the main positive (white/west) and negative (black/east) magnetic polarities at both ends. The AIA 9.4\,nm\ emission light curves of hot plasma near these regions show a similar evolution and both feet brighten almost simultaneously. This applies in particular to the initial, smaller brightenings visible in the light curves (see short vertical lines in panel (b)), which are clearly seen as transient bright dots in the accompanying online movie (distinguishable at the western footpoint from 11:50\,UT\ onward). The bulk of the loop shows similar evolution to the footpoints. The light curve from the AIA 17.1\,nm\ filter showing the evolution of cooler plasma (panel (d)) also displays high variability, but only near the eastern footpoint. Similar, high variability in the emission, although only at the eastern footpoint, is also apparent in the AIA 19.3\,nm\ and 21.1\,nm\ filter images. These intensity fluctuations near the eastern footpoint originate from a bright compact feature in the corona (see panel (c) and the movie). 

When these coronal emission maps are examined in conjunction with the photospheric magnetic field maps, we find that the bright compact coronal feature near the eastern footpoint lies directly above the region where the surface magnetic field is mixed with the presence of opposite magnetic polarities (see arrow in panel (e); magnetic field contours in panels (a) and (c)). Moreover, the surface magnetic field in the region of interest (black box in panel (e)) is evolving, in that we observe a monotonic decrease of the magnetic flux in the area at a rate of about $10^{15}$\,Mx\,s$^{-1}$ (panel (f)). We note that this decrease is due to the local disappearance of the magnetic flux and is not due to the advection of the flux out of the rectangular box considered. 


\begin{figure}
\begin{center}
\includegraphics[width=0.49\textwidth]{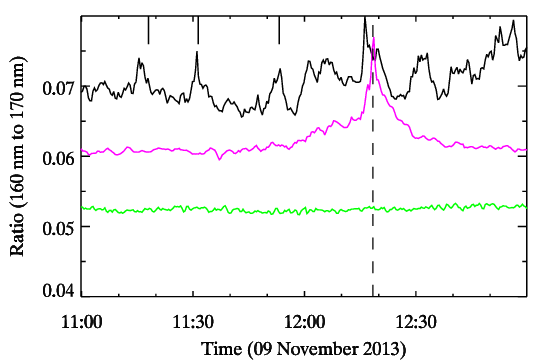}
\caption{As in Fig.\,\ref{fig:f1}(b) but for the ratios of 160\,nm and 170\,nm\ AIA filters. The green curve representing a quiet Sun region adjacent to the AR shows the comparison with the ratios from the footpoint regions (black and magenta, cf. Fig.\,\ref{fig:f1}). The short vertical bars and the vertical dashed line denote the same time-stamps as in Fig.\,\ref{fig:f1}(b). See Sect.\,\ref{sec:ar1} for details.
\label{fig:f2}}
\end{center}
\end{figure}

To examine the evolution of the lower atmosphere during the loop brightening, we adopt the ratio of AIA UV filters at 160\,nm\ and 170\,nm. The 160\,nm\ filter samples photospheric continuum and also has a contribution from the C\,{\sc iv}\ doublet forming at a characteristic transition region temperature of $\sim$0.1\,MK. The 170\,nm\ filter records photospheric continuum only \citep{2012SoPh..275...17L}. The intensity ratio of 160\,nm\ and 170\,nm\ filters can provide information on the C\,{\sc iv}\ emission, if present, by normalizing out continuum from both the filters. Therefore, we expect that the ratio of these light curves will be (i) close to constant if the C\,{\sc iv}\ doublet contribution to the 160\,nm\ filter is steady, and (ii) smaller compared to regions with higher densities of C\,{\sc iv}\ (meaning less plasma at 0.1\,MK assuming equilibrium conditions). In Fig.\,\ref{fig:f2} we plot such ratios from the eastern and western loop footpoints (black and magenta), and from a quiet-Sun region for reference (green). Unlike the almost time-independent (steady) and smaller (less plasma at 0.1\,MK) filter ratio from the quiet-Sun region, the ratios from the footpoint regions are akin to their coronal counterparts (particularly the 17.1\,nm\ light curves). The ratio from the eastern footpoint is similar to the intensity fluctuations observed in prolonged UV bursts, which are thought to result from magnetic reconnection in the chromosphere \citep[e.g.][]{2017A&A...605A..49C}.

Based on the presence of mixed polarities and the decrease in the magnetic flux at the surface, which coincides with the overlying brightenings near the eastern footpoint, we infer that this cancellation of magnetic flux is associated with magnetic reconnection. Here the magnetic energy released during the reconnection is likely responsible for the compact brightening, and the decrease in the surface magnetic flux is a consequence of the submergence of the reconnected field. We note that these mixed polarity magnetic concentrations are present only near the eastern footpoint but not at the western one (Fig.\,\ref{fig:f1}(e)). We cannot, however, rule out the presence of smaller and weaker patches of opposite polarity magnetic elements at the western footpoint due to the moderate spatial resolution of the HMI \citep[e.g.][cf. \citealt{2011SoPh..268....1B,2017ApJS..229....2S}]{2017ApJS..229....4C}.

The similarities in the AIA 9.4\,nm\ light curves from both the footpoints suggest that they are causally connected. Moreover, the compact bright region near the eastern footpoint is directly overlying a source, where the magnetic energy is likely released during the reconnection process. This indicates that the magnetic reconnection at the eastern footpoint is probably
also responsible for the disturbances seen at the western footpoint.

\section{Brightening in AR 12234}\label{sec:ar2}    


\begin{figure*}
\begin{center}
\includegraphics[width=\textwidth]{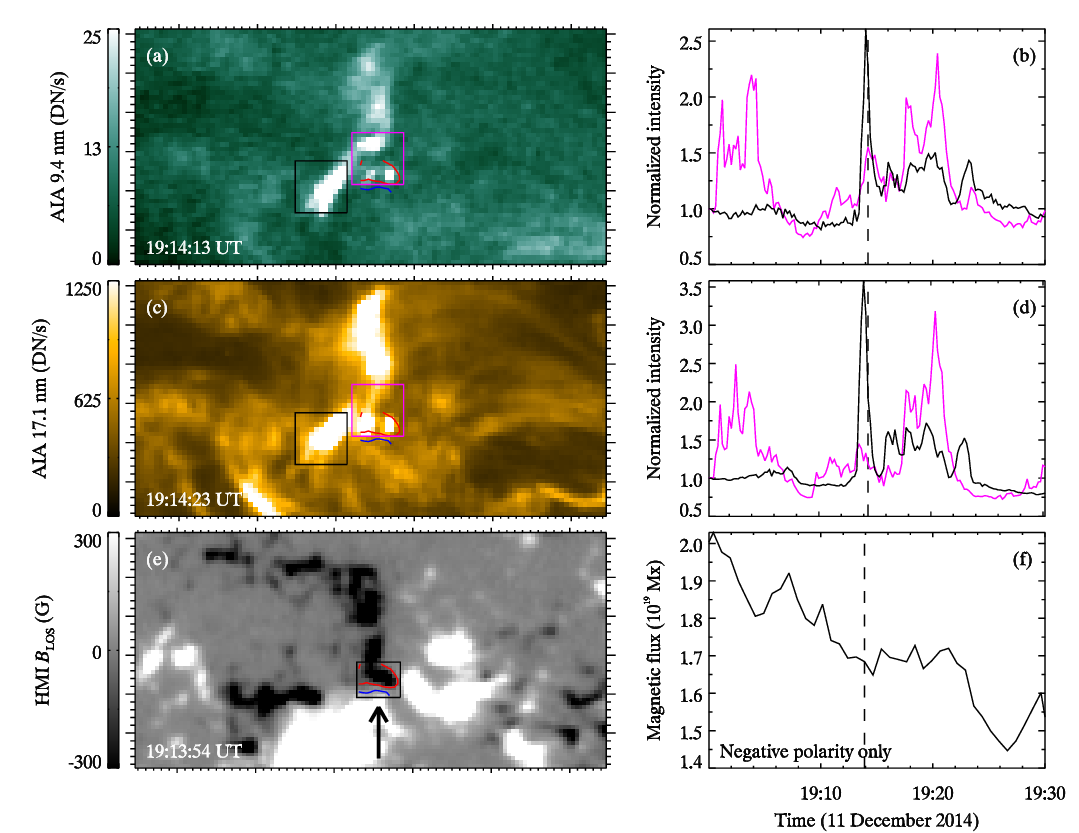}
\caption{As in Fig.\,\ref{fig:f1} but for AR 12234. The field of view is $60\arcsec\times30\arcsec$. North is up. See Sect.\,\ref{sec:ar2} for details.
\label{fig:f3}}
\end{center}
\end{figure*}

On 11 December 2014, during its second suborbital rocket flight, the Focusing Optics X-ray Solar Imager (FOXSI-2) detected a nanoflare in AR 12234, with faint emission in hard X-rays (above 3 keV), from plasma heated to temperatures above 10\,MK \citep{2017NatAs...1..771I}. The AR exhibited coronal brightenings recorded by AIA that are co-temporal to the inferred nanoflare in the same region. Unlike the AR-scale brightening displayed in Fig.\,\ref{fig:f1}, the AR 12234 brightening is a small-scale event extending about 15\arcsec\ (Fig.\,\ref{fig:f3}). Nonetheless, it shares similarities with the first example. 

The AR 12234 nanoflare is observed as a brightening in all the AIA EUV filters including the UV 160\,nm\ filter. In Fig.\,\ref{fig:f3}(a) we show the 9.4\,nm\ filter emission map during the reported nanoflare. The light curves from two different (east and west) regions covering the nanoflare exhibit large fluctuations (panel (b)). Clear, almost simultaneous peaks from both the regions at 19:14\,UT\ in the AIA 9.4\,nm\ filter are co-temporal with the hard X-ray emission detected by FOXSI\footnote{We note that being on a suborbital rocket flight, FOXSI-2 recorded data for about 6 minutes only and therefore the other peaks we display in Fig.\,\ref{fig:f3} occurred outside the window of time that  FOXSI could observe.} \citep{2017NatAs...1..771I}. For comparison, we show the AIA 17.1\,nm\ filter emission map and light curves that reveal similar characteristics to the 9.4\,nm\ filter (panels (c) and (d)).

The surface magnetic field underlying this nanoflare (panel (e)) is qualitatively similar to that near the east footpoint in Fig.\,\ref{fig:f1}. Here we also observe cancellation of the surface magnetic flux at a rate of about $10^{15}$\,Mx\,s$^{-1}$ (panel (f)). Therefore, also in the case of this FOXSI nanoflare the coronal brightening is causally associated with the dynamic evolution of the magnetic field in the form of flux cancellation at the surface.

\section{Energetics of coronal brightenings and flux cancellation events\label{sec:eng}}                               
Energy released in the process of magnetic reconnection during surface magnetic flux cancellation is generally considered to drive transient coronal brightenings, explosive events, and plasma jets in the upper solar atmosphere \citep[e.g.][]{1991JGR....96.9399D,1997Natur.386..811I,2008A&A...491..279C,2010PASJ...62..901M,2010ApJ...720.1136K,2014Sci...346C.315P,2014ApJ...795L..24T}. In both the cases discussed in Sections \ref{sec:ar1} and \ref{sec:ar2}, atmospheric brightenings associated with the loops are directly overlying regions of surface magnetic flux cancellation. Recent studies that employ magnetic field extrapolations locate the site of magnetic reconnection in the chromosphere (at heights of about 500\,km--1000\,km) in such flux cancellation events \citep[e.g.][]{2017A&A...605A..49C,2018ApJ...854..174T}. The location of the reconnection site can also be understood in terms of the horizontal separation at the solar surface between opposite-polarity magnetic concentrations while cancelling. Therefore, if the chromospheric reconnection is
also responsible for the observed loop brightenings, then, clearly, the exact energy input into the loops depends on the location of the reconnection, meaning that we can only provide an order-of-magnitude estimation. 

We use the loop brightening presented in Fig.\,\ref{fig:f1} as an example to estimate the energy content in the coronal section of the loop. The thermal energy density for a parcel of a mono-atomic gas at temperature, $T$, and particle density, $n$, is given by $e_\text{ther}=\frac{3}{2}nk_\text{B}T$, where $k_\text{B}$ is the Boltzmann constant. Here, we consider a typical coronal electron number density of $10^9$\,cm$^{-3}$, and a plasma temperature of 7\,MK,\ such that the loop is clearly visible in the AIA 9.4\,nm\ filter. The thermal energy of the loop is then $e_\text{ther}V$, where $V$ is the volume of the loop. From Fig.\,\ref{fig:f1}(a), the loop is roughly 10\,Mm\ wide and 100\,Mm long.\footnote{We estimate the loop length based on the lateral separation of footpoints and assuming it to be a semi-circle.} With these values, and the additional assumption that it has a circular cross-section, we estimate the thermal energy content of the loop to be $8\times10^{27}$\,erg. Next, we estimate the upper limit for the kinetic energy to be of the same order of magnitude as the thermal energy, because one can safely assume that the flows within the loop are subsonic (i.e. below about 400\,km\,s$^{-1}$). The energy required for the ionization of gas is negligible in comparison. Consequently, the total energy content of the loop is approximately $10^{28}$\,erg. 

Motivated by the close association of the flux cancellation with the loop brightening, we propose that the energy required for this loop brightening can be extracted from chromospheric reconnection at heights of around 500\,km.
Staying with the example from Fig.\,\ref{fig:f1}, we find that the cancelled flux amounts to about $\Phi=2.25\times10^{19}\,\text{Mx}$.
To estimate the converted energy in the flux cancellation, we use the (cancelled) magnetic energy $E_\text{mag}=\frac{1}{8\pi}B^2V$ in the volume $V=A\,H$ where the energy is released. We estimated the height range of energy release to be $H=500\,\text{km}$ above,\footnote{This estimate of $H$ is also consistent with three-dimensional MHD simulations of AR models, which yield a heating scale height below the corona of about 500\,km, i.e. in the chromosphere where the reconnection could be triggered \citep[i.e. the heating rate drops every 500 km by about a factor of 2.71; e.g.][]{2011A&A...530A.112B}.} and we estimate the area to be $A=3\,\text{Mm} \times 3\,\text{Mm}$ through the extent of the brightening near the eastern footpoint in Fig.\,\ref{fig:f1}(c).
The cancelled flux $\Phi=BA$ corresponds to the (cancelled) magnetic field $B$ integrated over the area $A$, meaning that we find the estimation for the magnetic energy converted during the flux cancellation to be given through $E_\text{mag}=\frac{1}{8\pi}\,\Phi^2H/A$.
With the above values, we find $E_\text{mag}=10^{28}\,\text{erg}$, which is comparable to the estimate for the energy content of the heated loop.
We note that this estimate for $E_\text{mag}$ might be a lower limit, because the reconnection region is 
most probably smaller, meaning that $A$ becomes smaller, and the ratio $H/A$ grows, resulting in a larger $E_\text{mag}$.
In principle, the magnetic energy released from chromospheric reconnection during flux cancellation is sufficient to explain the heating of the system of loops displayed in Fig.\,\ref{fig:f1}.

\section{Discussion}\label{sec:disc}    

Most of the magnetic energy from the flux cancellation events may already be dissipated low in the atmosphere, heating the chromosphere, where the energy requirements are even higher than in the corona \citep{1977ARA&A..15..363W}. However, if the observed coronal brightenings are indeed influenced by chromospheric (footpoint) reconnection during the flux cancellation (cf. Sect.\,\ref{sec:eng}), it is expected that some of the released magnetic energy reaches coronal heights. In this case, the issue of energy partition between the chromosphere and the corona needs to be addressed. Alternatively, the observed flux cancellation and footpoint reconnection can facilitate the release of the stored magnetic energy in coronal braids, leading to loop brightening.

Flux cancellation and associated reconnection can shed light on the spatial structuring (meaning only at specific locations in ARs) and temporal intermittency (meaning only at specific times) of hot loops observed in ARs \citep[e.g.][]{1981SoPh...69...99W,1992PASJ...44L.147S,1997ApJ...482..519F,2014ApJ...783...12U}. This is because such flux cancellation events, due to their sparsity, are discrete in space and time, at least on a scale visible to HMI. We note that the cancellation may in some cases be associated with previous small-scale flux emergence. Nevertheless, we emphasise that many flux cancellation events (and related reconnection) need not be associated with coronal brightenings due to the inherent lack of magnetic coupling to coronal heights at the locations of such events. For instance, some of the low-lying brightenings, such as UV bursts, which also occur over flux cancellation sites, do not show any coronal counterparts \citep[e.g.][]{2014Sci...346C.315P}.  

Another interesting aspect to be considered is the near-simultaneous brightening of two footpoints (which are laterally separated by about 100\arcsec) observed in the AIA 9.4\,nm\ filter (Fig.\,\ref{fig:f1}). Using the same observations, \citet{2014Sci...346B.315T} suggested that the brightening in the western footpoints and gentle upflows in the transition region are due to the acceleration of non-thermal particles in coronal nanoflares. Here we reveal ongoing magnetic reconnection at the eastern footpoint, presumably at low heights, that could accelerate non-thermal particles from the site of chromospheric reconnection. Further observational analysis is required to build statistics on the presence (or lack) of these near-simultaneous footpoint brightenings in association with brightenings of hot coronal loops \cite[e.g.][]{2018ApJ...857..137G}. In addition, it would be interesting to conduct numerical experiments to investigate the role of footpoint reconnection and the local energy deposition in the generation of a near-simultaneous signal at both footpoints. This is worthwhile because any energy transport mechanism (e.g. heat conduction or non-thermal particles) that is fast enough may efficiently transport the energy to the other footpoint within the AIA cadence of 12\,s, giving rise to an apparent near-simultaneous signal.  

In this study we focus on a scenario of coronal brightenings in AR cores and their likely association with chromospheric reconnection during surface magnetic flux cancellation. In principle, such a scenario might
also be applied to the diffuse solar corona in quiet Sun regions. The rate of energy loss from the entire quiet solar corona has been estimated to be about $10^{28}$\,erg\,s$^{-1}$ \citep{1977ARA&A..15..363W}. \citet{1988ApJ...330..474P} proposed nanoflares as a way to energize the quiet corona. With each nanoflare providing on average an energy of $10^{24}$\,erg, there would have to be about $10^4$ nanoflares at any given time distributed on the Sun to sustain the quiet corona.

In contrast to the nanoflares produced by reconnection distributed in the corona following energy build-up by braiding, one could also speculate that the quiet Sun is heated by small reconnection events at the coronal base associated with flux cancellation in the photosphere, similar to our proposal for the AR cores. In a recent study, \citet{2017ApJS..229...17S} used high-resolution magnetic field maps and analysed the rate of flux emergence and cancellation in the quiet Sun from the balloon-borne {\sc Sunrise} telescope \citep{2010ApJ...723L.127S,2011SoPh..268....1B}. For the magnetic features with a flux content in the range of 10$^{15}$\,Mx to 10$^{18}$\,Mx, they obtain a flux-loss rate (due to flux cancellation) per unit area of about $1150$\,Mx\,cm$^{-2}$\,day$^{-1}$ at the photosphere. This amounts to a flux-loss rate of $\dot{\Phi} = 8\times10^{20}$\,Mx\,s$^{-1}$, over the entire solar surface. This is a lower limit, as, using another technique, \citet{2013SoPh..283..273Z} obtain a flux-loss rate more than five times higher. 
We now consider that this flux loss leads to magnetic energy release through reconnection over a height range of $H=500$\,km above the photosphere, which corresponds to the temperature minimum level in traditional solar atmosphere models. Reconnection is most likely at these low levels as most of the small-scale loops in the quiet Sun do not reach much higher \citep{2010ApJ...723L.185W},
where the average magnetic field strength, $B$, is assumed to be in the range of 10\,G -- 100\,G\ \citep[e.g.][]{2004Natur.430..326T,2016A&A...593A..93D,2016A&A...596A...6L}. The rate of magnetic energy release due to flux cancellation can be written as $\dot{E} = \frac{1}{4\pi}{B\dot{\Phi}H}.$ The above values for $B$, $\dot{\Phi}$, and $H$ give a lower limit\footnote{We note that here we include the contribution of magnetic features with fluxes in the range of 10$^{15}$\,Mx -- 10$^{18}$\,Mx only. Therefore, our energy estimate should be considered as a lower limit.} for the rate of magnetic energy release of approximately $10^{28}-10^{29}$\,erg\,s$^{-1}$. Comparing this with the energy requirement of the quiet corona mentioned above shows that reconnection during small-scale flux cancellation would be sufficient to power the quiet corona. These order-of-magnitude estimates are consistent with previously reported values in the literature \citep[e.g.][]{2010SoPh..267...63Z}. If and how these small-scale reconnection events might be related to nanoflares is currently difficult to say due to the limitations in the detectability of an individual nanoflare.

\section{Conclusions}\label{sec:concl}  

We present two examples of recently reported nanoflares with an aim to understand their origin. To this end, we combined the diagnostics of atmospheric emission during the brightening of loops driven by nanoflares with the evolution of surface magnetic field. In both the cases we found that the surface magnetic field underlying the brightenings is dynamically evolving with clear signatures of a monotonic decrease in the magnetic flux (at a rate of about $10^{15}\,\text{Mx s}^{-1}$), suggesting the release of magnetic energy through reconnection at one of the apparent footpoints of the loops hosting the nanoflares. The surface evolution of the magnetic field connecting the coronal brightenings highlights a complex and dynamic magnetic coupling through the solar atmosphere that probably governs the energetics of these coronal structures and, as we speculate, those of many others. We suggest that the energy released in chromospheric reconnection is a viable source to power AR coronal loops driven by nanoflares. A similar process of chromospheric reconnection during flux cancellation would be sufficient to power the quiet solar corona.

\begin{acknowledgements}
We thank the anonymous reviewer for constructive comments, which helped to improve the manuscript. L.P.C. benefited from many useful discussions with Eric Priest, and received funding from the European Union's Horizon 2020 research and innovation programme under the Marie Sk\l{}odowska-Curie grant agreement No.\,707837. This project has received funding from the European Research Council (ERC) under the European Union's Horizon 2020 research and innovation programme (grant agreement No.\,695075). This work was partly supported by the BK21 plus program through the National Research Foundation (NRF) funded by the Ministry of Education of Korea. SDO data are courtesy of NASA/SDO and the AIA, and HMI science teams. This research has made use of NASA's Astrophysics Data System.
\end{acknowledgements}



\end{document}